\documentclass[10pt,a4paper,twocolumn,english,prl,aps,showpacs,floatfix,groupedaddress,superscriptaddress]{revtex4}
\usepackage[]{fontenc}
\usepackage[latin1]{inputenc}
\usepackage{verbatim}
\usepackage{amsmath}
\usepackage{graphicx}
\usepackage{amssymb}

\makeatletter

\def\1{1\negthickspace{\rm I}}

\usepackage{babel}
\makeatother
\begin{document}

\title{Qubit Teleportation and Transfer across Antiferromagnetic Spin Chains}

\author{L. Campos Venuti}

\affiliation{Institute for Scientific Interchange (ISI), Villa Gualino, viale
Settimio Severo 65, I-10133 Torino, Italy}

\affiliation{INFN Sezione di Bologna, viale C. Berti-Pichat 6/2, I-40127 Bologna,
Italy}

\author{C. Degli Esposti Boschi}

\affiliation{CNR, Unità CNISM di Bologna, viale C. Berti-Pichat 6/2, I-40127 Bologna,
Italy}

\affiliation{Dipartimento di Fisica, Università di Bologna, viale C. Berti-Pichat
6/2, I-40127 Bologna, Italy}

\author{M. Roncaglia}

\affiliation{Dipartimento di Fisica, Università di Bologna, viale C. Berti-Pichat
6/2, I-40127 Bologna, Italy}

\affiliation{INFN Sezione di Bologna, viale C. Berti-Pichat 6/2, I-40127 Bologna,
Italy}

\date{\today}

\begin{abstract}
We explore the capability of spin-$1/2$ chains to act as quantum
channels for both teleportation and transfer of qubits. Exploiting
the emergence of long-distance entanglement in low-dimensional systems
{[}Phys. Rev. Lett. \textbf{96}, 247206 (2006)], here we show how
to obtain high communication fidelities between distant parties. An
investigation of protocols of teleportation and state transfer is
presented, in the realistic situation where temperature is included.
Basing our setup on antiferromagnetic rotationally invariant systems,
both protocols are represented by pure depolarizing channels. We propose
a scheme where channel fidelity close to one can be achieved on very
long chains at moderately small temperature.
\end{abstract}


\maketitle
%

\emph{Introduction.} In order to accomplish the main tasks of Quantum
Information, a sizable amount of entanglement is needed \cite{nielsen&chuang2000}.
In addition, the particles that share entanglement must be accessed
individually for measurements and, quite importantly, they must be
well separated in space. 

Recently it was shown \cite{campos06LDE} that in some spin models
at zero temperature (i.e.~in the ground state) a selected pair of
distant sites $A$ and $B$ can be highly entangled. In some cases
sites $A$ and $B$ may be taken infinitely far apart still retaining
a high amount of entanglement, a situation that was termed long distance
entanglement (LDE). An example of this situation is given by the end-sites
of an open $S=1/2$ dimerized Heisenberg chain. Even for moderate
values of the dimerization this effect is strong enough to develop
non local correlations, i.e.~entanglement, between the end-sites
of an open chain of infinite length. 

The main aim of this Letter is to explore the actual feasibility of
quantum teleportation and transfer across spin $1/2$ chains that
exhibit LDE. Having in mind realistic optical lattice implementations
of spin chains \cite{hofstetter06}, we consider the principal cause
of decoherence which is given by the temperature. Using the same schemes
proposed in Ref. \cite{campos06LDE}, we expect the entanglement between
$A$ and $B$ to deteriorate when the temperature becomes of the order
of the lowest excitation gap $\Delta$. As this gap, which originates
from the boundary conditions, typically vanishes when the length $L$
of the chain increases, we are led to explore the tradeoff between
temperature and chain length. 

As will be clarified throughout this paper, antiferromagnetic chains
with global SU(2)-invariance have several advantages. Typically, in
these systems rotational symmetry is never broken. As a consequence
the two-particle reduced density matrix $\rho_{AB}$ (obtained by
tracing the total $\rho$ over all the Hilbert space except sites
$A$ and $B$) maintains SU(2) invariance, i.e.~it is a Werner state
\cite{werner89} in the language of quantum information. Werner states
are described by a single parameter which can be taken to be $\langle\sigma_{A}^{z}\sigma_{B}^{z}\rangle_{\rho}=\mathrm{Tr}\left(\rho_{AB}\sigma_{A}^{z}\sigma_{B}^{z}\right)\in\left[-1,1/3\right]$.
The interval $\langle\sigma_{A}^{z}\sigma_{B}^{z}\rangle_{\rho}\in[-1,-1/3)$
corresponds to entangled $\rho_{AB}$. 

At $T=0$ the density matrix is $\rho=|G\rangle\langle G|$, with
$|G\rangle$ the ground state, while at finite temperature it is given
by the canonical density operator $\rho=Z^{-1}e^{-\beta H}$, with
$\beta=1/T$ (in units of $k_{B}$) and $Z$ the normalization factor.
At low temperatures we can approximate the thermal density matrix
by retaining only the ground state and the first excited states. On
quite general grounds \cite{lieb62} the ground state $|G\rangle$
is a total singlet, while the first excitations are given by a spin
one triplet $|m\rangle$ labeled by the total magnetization: $S_{\mathrm{tot}}^{z}=m=-1,0,1$.
Then at low temperatures we can write\begin{equation}
e^{-\beta H}\simeq e^{-\beta E_{0}}\left[|G\rangle\langle G|+e^{-\beta\Delta}\sum_{m=-1,0,1}|m\rangle\langle m|\right],\label{eq:thermal}\end{equation}
where $E_{0}$ is the ground state energy and $\Delta$ is the first
excitation gap. Notice that this approximation correctly maintains
rotational invariance. The thermal reduced density matrix $\rho_{AB}\left(T\right)$
of $A$ and $B$ depends only on the following average value\begin{align}
\langle\sigma_{A}^{z}\sigma_{B}^{z}\rangle_{T} & =\frac{1}{1+3e^{-\beta\Delta}}\left[\langle G|\sigma_{A}^{z}\sigma_{B}^{z}|G\rangle{{\atop }}\right.\nonumber \\
 & \left.{{\atop }}+e^{-\beta\Delta}\left(\langle1|\sigma_{A}^{z}\sigma_{B}^{z}|1\rangle+2\langle1|\sigma_{A}^{x}\sigma_{B}^{x}|1\rangle\right)\right],\label{eq:szsz_T}\end{align}
which has been written exploiting the SU(2) invariance. The form (\ref{eq:szsz_T})
is particularly useful for numerical density-matrix renormalization-group
(DMRG) simulations \cite{schollwock05} since it involves only the
computation of the lowest-state correlation functions in the sectors
$m=0$ and $m=1$. 

In the situations analyzed in \cite{campos06LDE} where LDE is present
in the ground state, the $S=1$ triplet state is localized near the
sites $A$ and $B$. As we will show below, the entanglement in $\rho_{AB}\left(T\right)$
is maintained until $T$ becomes comparable with the gap $\Delta$,
when the triplet state becomes non-negligible. We are then led to
prefer systems with a large gap $\Delta$. Quite generally however,
open systems with a finite bulk correlation length $\xi$ develop
mid-gap levels scaling exponentially with the system size $\Delta\simeq e^{-L/\xi}$
\cite{white&huse93}. On the other hand, systems with a diverging
correlation length give rise to an algebraic decay, $\Delta\sim L^{-\alpha}$.
The generality of this conjecture -- that establishes a relation between
bulk correlation length and the decay of the mid-gap -- is a challenging
question that deserves further studies. 

\begin{figure}[h]
\begin{centering}\includegraphics[width=6cm]{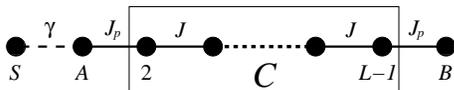}\par\end{centering}

\caption{Model Hamiltonian considered for teleportation (joint measure between
$S$ and $A$) and for transfer (switching on $\gamma$ at a given
time).\label{fig:model}}
\end{figure}

\begin{figure}
\begin{centering}\includegraphics[scale=0.35]{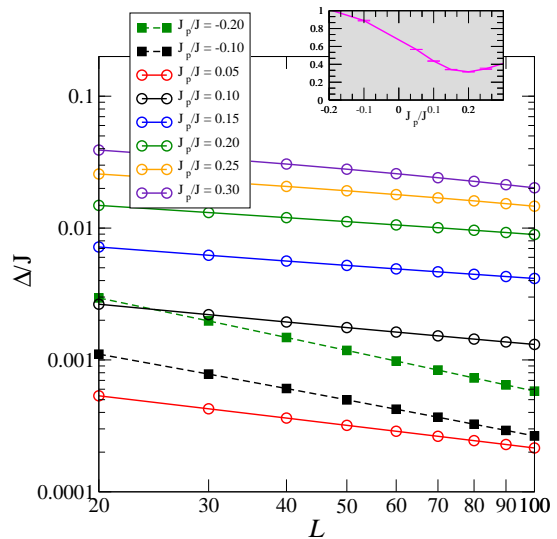}\par\end{centering}

\caption{Finite size scaling behavior of the lowest gaps. In the inset is
plotted the scaling exponent fitted with the law $\Delta=cL^{-\alpha}$.
The data were obtained with a DMRG code using 400-500 optimized states
and three finite system sweeps. \label{fig:scaling_gaps}}
\end{figure}

For the above-mentioned reasons we propose to use an open $S=1/2$
Heisenberg chain with different interactions at the endpoints \begin{equation}
H_{\mathrm{chain}}=H_{C}+J_{p}\left(\vec{S}_{A}\cdot\vec{S}_{2}+\vec{S}_{B}\cdot\vec{S}_{L-1}\right),\label{eq:H_chain}\end{equation}
as depicted in Fig.~\ref{fig:model} (system $ACB$). In such a system
there is strictly no LDE in the thermodynamic limit, but for finite
size one can always choose $J_{p}/J$ small enough so as to have arbitrarily
large entanglement between $A$ and $B$ in the ground state. Moreover
we checked numerically that in this system the first gap $\Delta$
scales only algebraically with the size of the system $L$: $\Delta\sim L^{-\alpha}$
as can be seen in Fig.~\ref{fig:scaling_gaps}. Note the slow decay
of the gaps due to the small value of $\alpha$ (see inset). %

\emph{Teleportation.} Entangled Werner (SU(2) invariant) states have
several advantages when used as a resource for quantum informational
devices. As far as teleportation is concerned, one can show \cite{horodeckis99}
that the standard teleportation scheme \cite{bennett1993tele} is
the best over all possible schemes at least in the region where a
better-than-classical fidelity is achieved. In the standard protocol
an unknown state $\xi$ at site $S$ (see Fig. \ref{fig:model}) is
teleported to site $B$ by making a joint Bell measurement on sites
$S$ and $A$ and transmitting the result of the measurement $j$
to $B$ where a unitary transformation is applied. If $A$ and $B$
share a pure maximally entangled (SU(2) invariant) state $|\psi^{-}\rangle_{AB}=(\left|\uparrow\downarrow\right\rangle _{AB}-\left|\downarrow\uparrow\right\rangle _{AB})/\sqrt{2}$,
then the state $\xi$ is transferred to $B$ exactly. In a realistic
situation, external noise of any kind turns the pure state $|\psi^{-}\rangle_{AB}$
into a non maximally entangled mixed state $\rho_{AB}$. %

Using this protocol with a Werner state as resource, the fidelity
of teleportation does not depend on the the outcome $j$ nor on the
state to be teleported. By repeating the experiment many times with
the same input state, the teleportation process is represented by
a quantum channel mapping input states $\xi$ at site $S$ into teleported
states $\Lambda\left(\xi\right)$ at site $B$ \cite{bowen2001}.
In this case, the teleportation channel is given precisely by a pure
depolarizing channel:\begin{equation}
\Lambda\left(\xi\right)=\vartheta\xi+\left(1-\vartheta\right)\frac{1}{2}\mathbb{I}.\label{eq:tele_depol}\end{equation}
The parameter $\vartheta$ which identifies the channel -- sometimes
called shrinking factor -- takes the simple form $\vartheta=-\langle\sigma_{A}^{z}\sigma_{B}^{z}\rangle$.
Obviously, $\Lambda$ turns into an ideal channel when $\vartheta=1$,
i.e.~when $\rho_{AB}$ is the singlet $|\psi^{-}\rangle_{AB}$. %
 The fidelity of teleportation is \[
f=\mathrm{Tr}\left(\xi\Lambda\left(\xi\right)\right)=\frac{1+\vartheta}{2}=\frac{1-\langle\sigma_{A}^{z}\sigma_{B}^{z}\rangle}{2},\]
that indeed does not depend on the state to teleport.

\begin{figure}
\begin{centering}\includegraphics[scale=0.4]{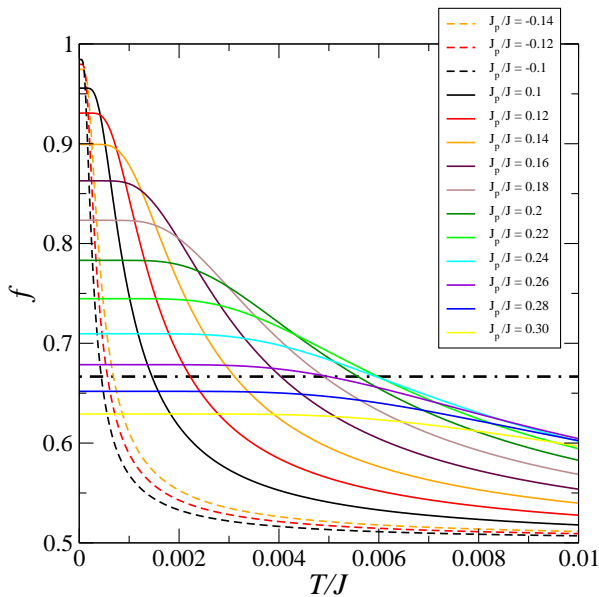}\par\end{centering}

\caption{Fidelity of teleportation between end sites $A$ and $B$ as a function
of temperature for the Heisenberg model. The curves refer to different
values of the interaction $J_{p}$. \label{fig:Fidelity-of-teleportation}}
\end{figure}

For our class of systems, $\vartheta$ is given by Eq.~(\ref{eq:szsz_T}).
When the temperature is increased from zero, it eventually reaches
a value $T^{\ast}$, above which the thermal state $\rho_{AB}\left(T\right)$
becomes separable. This occurs when $\langle\sigma_{A}^{z}\sigma_{B}^{z}\rangle_{T^{\ast}}=-1/3$,
that gives\begin{equation}
T^{*}=\Delta\left[\log\left(\frac{\langle1|\sigma_{A}^{z}\sigma_{B}^{z}|1\rangle+2\langle1|\sigma_{A}^{x}\sigma_{B}^{x}|1\rangle+1}{-\langle G|\sigma_{A}^{z}\sigma_{B}^{z}|G\rangle-1/3}\right)\right]^{-1}.\label{eq:T-threshold}\end{equation}
Typical values in our scheme are obtained using the two qubit singlet
and triplet pure states, for which we get $T^{*}=\Delta/\log\left(3\right)\approx0.9\Delta$.
The gap $\Delta$ and the correlations appearing in Eq.~(\ref{eq:T-threshold})
can be calculated numerically as functions of $L$ and $J_{p}$. In
Fig.~\ref{fig:Fidelity-of-teleportation} we plot the results, obtained
from DMRG simulations for a chain of $L=50$ sites.%
 In view of an optical lattice experiment, these curves could serve
to locate the working point to achieve the maximal possible fidelity. 

\emph{State transfer.} As suggested by S. Bose \cite{bose03} open
spin chains can be exploited for transferring quantum states from
one end to the other end of the chain. Let a chain of length $L$
be described by the Hamiltonian $H_{\mathrm{chain}}$. For times $t<0$
the chain is in its ground state or possibly in a state of thermal
equilibrium $\rho_{\mathrm{chain}}=Z^{-1}e^{-\beta H_{\mathrm{chain}}}$.
At a time $t=0$ a spin-spin interaction $\gamma\vec{S}_{S}\cdot\vec{S}_{A}$
between the sender $S$ (that stores the pure state to be transferred
$|\xi\rangle$) and site $A$ is switched on and let to evolve with
the Hamiltonian, as depicted in Fig. \ref{fig:model}\begin{equation}
H=H_{\mathrm{chain}}+\gamma\vec{S}_{S}\cdot\vec{S}_{A}.\label{eq:ham_tot}\end{equation}
%
After a given optimal time $t^{\ast}$, the initial state $|\xi\rangle$
gets transferred to site $B$ with fidelity $f$. 

We stress here the importance of dealing with antiferromagnetic interactions.
In this case, elementary excitations typically have relativistic linear
dispersion for small momenta, i.e.~$\omega\left(k\right)\simeq v\left|k\right|$
where $v$ is the effective speed of light. On the contrary, in ferromagnetic
systems, as the one originally proposed in \cite{bose03}, the dispersion
of elementary excitations is generally quadratic for small momenta.
This fact leads to dispersive effects which limit the fidelity of
transfer. 

From a quantum information perspective, one can easily show that the
state transfer protocol with SU(2) invariant systems is precisely
given by the depolarizing channel given by (\ref{eq:tele_depol}).
The unique parameter specifying the channel is given in this case
by $\vartheta=\langle\sigma_{B}^{z}\left(t\right)\rangle_{\rho}$,
$\rho=\left|\uparrow\right\rangle \left\langle \uparrow\right|\otimes\rho_{\mathrm{chain}}$,
where the time evolution is according to the total Hamiltonian (\ref{eq:ham_tot}).
%
 The calculation of this quantity in a strongly correlated system
is a non-trivial task. However, an approximation scheme is possible
for the models where we observed LDE (or quasi-LDE). 

Although the spins on $A$ and $B$ do not interact directly, they
experience an effective interaction mediated by the system $C$. Due
to rotational invariance, the model (\ref{eq:H_chain}) is effectively
mapped, at every perturbative order, onto an SU(2)-symmetric Hamiltonian
for the sites $A$ and $B$ \begin{equation}
H_{\mathrm{eff}}=J_{\mathrm{eff}}\vec{S}_{A}\cdot\vec{S}_{B},\label{eq:H_eff}\end{equation}
%
 This approximation holds when the energy splitting $J_{\mathrm{eff}}$
caused by $H_{\mathrm{eff}}$ is smaller than the typical gaps in
the unperturbed Hamiltonian $H_{C}$. On the one hand, we know from
conformal field theory \cite{henkel99} that finite-size gaps in $H_{C}$
scale as $JL^{-1}$. On the other hand, $J_{\mathrm{eff}}$ is nothing
but the singlet-triplet gap $\Delta$. We have numerically checked
that $\Delta$ scales as $JL^{-\alpha}$, in the system $H_{\mathrm{chain}}$,
as can be seen from Fig.~\ref{fig:scaling_gaps}. The correct prefactor
has the form $\Phi(J_{p}/J)$. From perturbation theory we know that
$\Phi(x)\approx x^{2}$ for small $x$. This means that we can reliably
approximate the model (\ref{eq:H_chain}) with the effective Hamiltonian
(\ref{eq:H_eff}), provided that $\Phi(J_{p}/J)<L^{\alpha-1}$, i.e.
$J_{p}<JL^{(\alpha-1)/2}$ when $J_{p}$ is small enough. 

Our scheme of approximation reduces the state transfer protocol to
an effective three site problem where the time evolution is unitary
by means of the Hamiltonian $H=\gamma\vec{S}_{S}\cdot\vec{S}_{A}+J_{\mathrm{eff}}\vec{S}_{A}\cdot\vec{S}_{B}$.
The average is done with respect to the ensemble $\rho_{0}=|\xi\rangle\langle\xi|_{S}\otimes\rho_{AB}$,
where $\rho_{AB}$ is the most general mixed state which preserve
SU(2) invariance, i.e.\[
\rho_{AB}=\frac{1}{4}\mathbb{I}+\frac{g}{4}\vec{\sigma}_{A}\cdot\vec{\sigma}_{B},\]
and $g=\langle\sigma_{A}^{z}\sigma_{B}^{z}\rangle$ that includes
decoherence effects from the environment $C$ as well as the effect
of temperature. 

\begin{figure}[t]
\begin{centering}\includegraphics[scale=0.4]{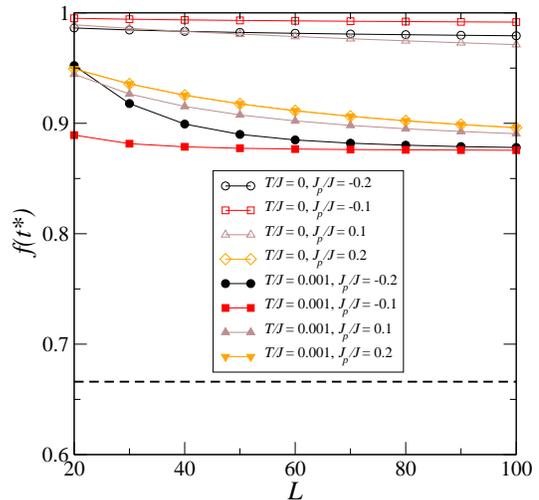}\par\end{centering}

\caption{Transfer fidelity at optimal time as a function of the chain length
$L$. The curves refer to some values the coupling $J_{p}$ between
the probes ($A$ and $B$) and the Heisenberg chain. Results are reported
at both zero and finite temperature. \label{fig:transfer_fidelity}}
\end{figure}

The density matrix evolves as\[
\rho\left(t\right)=e^{-itH}\left(|\xi\rangle\langle\xi|_{S}\otimes\rho_{AB}\right)e^{itH}.\]
 The fidelity of the transfer from site $S$ to site $B$ at a given
time $t$ is $f\left(t\right)=\mathrm{Tr}_{SAB}\left(\rho\left(t\right)|\xi\rangle\langle\xi|_{B}\right)$.
After some calculations we get\begin{eqnarray}
f\left(t\right) & = & \frac{1}{36\omega^{2}}\left\{ \left(22+4g\right)\left(J_{\mathrm{eff}}^{2}+\gamma^{2}\right)-\gamma J_{\mathrm{eff}}\left(19+10g\right)\right.\nonumber \\
 & - & 2\left(1+g\right)\omega\left[\omega_{-}\cos\left(t\omega_{+}/2\right)+\omega_{+}\cos\left(t\omega_{-}/2\right)\right]\nonumber \\
 & + & \left.3\gamma J_{\mathrm{eff}}\left(2g-1\right)\cos\left(\omega t\right)\right\} ,\label{eq:fidel}\end{eqnarray}
where $\omega_{\pm}=\omega\pm(J_{\mathrm{eff}}+\gamma)$ and $\omega=\sqrt{J_{\mathrm{eff}}^{2}-J_{\mathrm{eff}}\gamma+\gamma^{2}}$.
The maximal possible interference (constructive and destructive) is
achieved when the frequencies are commensurate each other i.e.~for
$\gamma=J_{\mathrm{eff}}$. In this case the fidelity reduces to \begin{align*}
f\left(t\right) & =\frac{1}{36}[25-2g-6\left(1+g\right)\cos\left(J_{\mathrm{eff}}t/2\right)\\
 & +\left(6g-3\right)\cos\left(J_{\mathrm{eff}}t\right)+2\left(1+g\right)\cos\left(3J_{\mathrm{eff}}t/2\right)].\end{align*}
The first maximum of this function is attained at a time \begin{eqnarray}
t^{\ast} & = & \frac{2}{J_{\mathrm{eff}}}\arccos\left(\frac{1-2g-\sqrt{12g^{2}+12g+9}}{4\left(1+g\right)}\right)\nonumber \\
 & = & \frac{\pi}{J_{\mathrm{eff}}}+\frac{2}{3}\frac{\left(g+1\right)}{J_{\mathrm{eff}}}+O\left(\left(g+1\right)^{2}\right).\label{eq:tstar}\end{eqnarray}
The value $g=-1$ represents the ideal case where we have a pure singlet
$\rho_{AB}=|\psi^{-}\rangle\langle\psi^{-}|_{AB}$ at our disposal,
with $t^{\ast}=\pi/\omega$. In the non-ideal case, the time for best
transfer gets only slightly shifted by a value which in the worst
case ($g=1/3$) is $1.448$. The maximum fidelity is then \begin{align}
f^{\ast} & =f\left(t^{\ast}\right)=\frac{\sqrt{3}\left(4g^{2}+4g+3\right)^{3/2}+24g^{2}+66g+33}{48\left(1+g\right)^{2}}\nonumber \\
 & =1-\frac{2}{9}\left(g+1\right)+\frac{1}{18}\left(g+1\right)^{2}+O\left(\left(g+1\right)^{3}\right).\label{eq:fstar}\end{align}
As expected, the transfer is perfect for $g=-1$. However, the transfer
fidelity remains very high for all the possible values $g\in\left[-1,1/3\right]$.
The lowest possible value $f^{\ast}=7/8$ is attained at $g=0$ (maximally
mixed case). Anyway, we must restrict ourself to the situation where
the approximation of unitary evolution is valid, i.e. $g\simeq-1$,
and correspondingly the transmission fidelity is very close to 1.
As in the teleportation case, we have considered the transfer of a
state using a Heisenberg chain playing the role of system $C$. In
Fig.~\ref{fig:transfer_fidelity} the optimal transfer fidelity is
plotted as a function of the chain length $L$ at temperature $T=0$
and $T=10^{-3}J$, for some values of $J_{p}$. In any case, we find
a more-than-classical transmission fidelity even for chains of length
100 sites. For obtaining these results, the existence of entanglement
between the two distant sites $A$ and $B$ is crucial . %

Now, let us draw an additional consideration. First, we note that
$t^{\ast}\propto J_{\mathrm{eff}}^{-1}=L^{\alpha}/(J\Phi)$, with
$\alpha<1$ (see Fig.~\ref{fig:scaling_gaps}) On the other hand,
our scheme is expected to be valid under the condition $\Phi(J_{p}/J)\lesssim L^{\alpha-1}$
which implies $t^{\ast}\gtrsim L/J$, consistently with the {}``flying''
qubit picture where the information is carried by elementary spin
excitations. 

Finally, we mention that the transfer protocol may be used also for
sharing entanglement between distant parties \cite{bose03}. In our
situation, we already have entanglement between distant parties $A$
and $B$, but we can ask how it may be further increased. 

The idea is to start having a maximally entangled singlet $\rho_{\mathrm{in}}=|\psi^{-}\rangle\langle\psi^{-}|_{XS}$
at sites $S$ and at an extra neighboring site $X$ completely decoupled
from the rest. Then, we send the $S$ part of the input state $\rho_{\mathrm{in}}$
through the quantum channel described by our transmission protocol.
At a certain time $t^{\ast}$, we obtain an outcome state living on
the pair of sites $X$ and $B$ \[
\rho_{\mathrm{out}}=\left(1-p\right)\rho_{\mathrm{in}}+\frac{p}{3}\sum_{k=1}^{3}\mathbb{I}\otimes\sigma^{k}\rho_{\mathrm{in}}\mathbb{I}\otimes\sigma^{k},\]
where $p=3\left(1-\vartheta\right)/4$ is the so called error probability.
In this state, the concurrence between $X$ and $B$ is \begin{equation}
C\left(\rho_{\mathrm{out}}\right)=\mbox{max}\left[1-2p,0\right]=\max\left[3f^{\ast}-2,0\right],\label{eq:conc_new}\end{equation}
while original state $\rho_{AB}$ had a concurrence given by $C\left(\rho_{AB}\right)=\mbox{max}\left[-3/2g-1/2,0\right]$.
Using $f^{\ast}$ from Eq.~(\ref{eq:fstar}) it is possible to estimate
that the concurrence is increased, i.e. $C\left(\rho_{\mathrm{out}}\right)\geq C\left(\rho_{AB}\right)$,
where the equality holds only when $g=-1$ (it is not possible to
increase the entanglement of a singlet). The minimum value is achieved
for the completely mixed case $g=0$ where the concurrence is $C\left(\rho_{\mathrm{out}}\right)=5/8=0.625$.

\emph{Conclusions.} We have given an explicit evidence that open antiferromagnetic
Heisenberg chains may represent good quantum channels for teleportation
and state transfer. This result relies mainly on the possibility to
entangle the two end-spins (quasi-LDE) by choosing an appropriate
coupling $J_{p}$. We have shown that, despite the smallness of the
lowest gap, high fidelities of both teleportation and transfer may
be achieved, with a tradeoff between temperature and chain length.
It is tempting to speculate about the possibility of reproducing these
effects in optical lattice environments \cite{campos_fut_illu}.  

We thank J.I. Cirac, M. Giampaolo, F. Illuminati, M. Keyl, D. Porras
for interesting discussions and G. Morandi for a careful reading of
the manuscript. This work was partially supported by the COFIN projects
2002024522\_001 and 2003029498\_013. 

\bibliographystyle{apsrev}
\bibliography{tele_LDE}

\end{document}